\documentclass[12pt]{article}

\usepackage{epsfig,amsmath,amsfonts,amssymb}

\usepackage{bm}

\newcommand{\beq}{\begin{equation}}
\newcommand{\eeq}{\end{equation}}
\newcommand{\beqar}{\begin{eqnarray}}
\newcommand{\eeqar}{\end{eqnarray}}
\newcommand{\pa}{\partial}

\newcommand{\IV}{$I$-$V$}
\newcommand{\etal}{\textit{et al}.}

\newcommand{\refart}[7]{{#1}, {#3} \textbf{#4}, {#5} {(#6)}}%

\newcommand{\refbook}[5]{{#1}, \textit{#2} (#3, #4, #5)}%

\newcommand{\prb}{Phys.\,Rev.\,B}
\newcommand{\prd}{Phys.\,Rev.\,D}
\newcommand{\prl}{Phys.\,Rev.\,Lett.}

\begin{document}

\title{SQUID-based Resonant Detection \\of Axion Dark Matter }
\author{Vladimir A. Popov}%
\date{}
\maketitle%
%\email{vladipopov@mail.ru}
%\affiliation{
\begin{center}{\small Institute of Physics, Kazan Federal University, \\Kremlyovskaya st.~18, Kazan 420008, Russia}\end{center}

\vspace{1cm}

\begin{abstract}%must be specified before the \maketitle command
A new method for searching for Dark Matter axions is proposed. It is shown that a two-contact SQUID can detect oscillating magnetic perturbations induced by the axions in a strong inhomogeneous magnetic field. A resonant signal is a steplike response in the SQUID current-voltage characteristic at a voltage corresponding to the axion mass with a height depending on the axion energy density near the Earth. The proposed experimental technique appears to be sensitive to the axions with masses $m_a\lesssim 10^{-4}$ eV, which is well-motivated by current researches both in cosmology and in particle physics.
\end{abstract}

%\pacs{14.80.Va, 85.25.Dq, 95.35.+d}
%\keywords{axion, dark matter, SQUID, Josephson junction}

\vspace{1cm}

To understand the nature of the Dark Matter (DM) is among major challenges in the present-day cosmology.  A number of particles is considered as DM candidates (WIMPs, sterile neutrinos, ets.) and low mass axions are highly attractive ones. The experimental discovery of the axions would give new insights into cosmological and astrophysical researches as well as into particle physics since the axions play a central role in the solution to  the strong $CP$-problem. This provides that experimental searching for the axions with mass in the range of $10^{-6}-10^{-3}$ eV is of paramount importance.

The experimental techniques \cite{ADMX,throuWall,ALPS,CAST} for DM axions detection are based on axion-phonon conversation processes. Their theoretical description is developed in the conventional manner for extensions of the standard model of particle physics from the term $-\frac{1}{4}gaF_{\alpha\beta}\tilde F^{\alpha\beta}$ in the Lagrangian density \cite{Sikivie1}.

A peculiar approach to axion detection using a Josephson junction (JJ) was proposed in the recent letter \cite{Beck1}. It is based on a hypothesis that a phase difference in the JJ and an axion misalignment angle $\theta$ are related to each other. This means that the axions directly govern the supercurrent across the junction, $I=I_\text{c}\sin\theta$. This assumption also implies a more fundamental physical implication that there exists a quantum interference between the DM axions \cite{comment1}, which possibly form a cosmic Bose-Einstein condensate \cite{Sikivie2}, and Cooper pairs.  A possible experimental corroboration for this approach is a resonant peak of unknown origin available in Ref.~\cite{Hoffmann} and less evident signals collected in \cite{Beck2}. There is no doubt that this result needs further comprehensive verifications to exclude other reasons for the signal such as subharmonic Shapiro steps \cite{Belenov}, and thus to be sure in its axionic origin.

Using superconducting quantum interference devices (SQUIDs) is a more usual way to employ the JJs in the axion searching experiments. To use the SQUIDs inevitably comes in mind because an expected response from the axions is very weak and a high sensitive devices are required to detect it. During the last year several new experimental techniques, which use  SQUIDs as  magnetometers, were suggested \cite{Graham,Budker,Sikivie3}.

In this letter we consider an alternative possibility for galactic halo axion detection by application of SQUIDs.  The suggested approach exploits resonant properties of the JJs and follows the conventional notion of JJs and of axions and their interactions with ordinary particles. The corresponding effective Lagrangian for the axion-photon system is (we use natural units, $\hbar=c=1$)
\beq\label{lagrangian}
{\cal L} = -\frac{1}{4}F_{\alpha\beta}F^{\alpha\beta} + \frac{1}{2}\left(\pa_\alpha\pa^\alpha a - m_a^2 a^2\right) - \frac{1}{4}gaF_{\alpha\beta}\tilde F^{\alpha\beta},
\eeq
where $a$ is the axion field and $m_a$ is its mass, $F_{\alpha\beta}$ is the electromagnetic field tensor and $\tilde F^{\alpha\beta}$ is its dual. The third term describes the $CP$-invariant interaction between the pseudoscalar and electromagnetic fields. It inevitably comes about when the Peccei-Quinn symmetry is spontaneously broken at energy scales of axion decay constant $f_a$. The coupling constant $g=g_\gamma\alpha/\pi f_a$, where $\alpha$ is the fine-structure constant and $g_\gamma$ is a dimensionless model-dependent parameter. Its value is $g_\gamma=0.97$ for the KSVZ model \cite{Kim,Shifman} and $g_\gamma=-0.36$ for the DFSZ model \cite{Zhitnitskii,Dine}.
The Peccei-Quinn mechanism provides that the product of the axion mass and the decay constant is of order of the same product for pions $m_af_a\approx \frac{1}{2}m_\pi f_\pi \approx 6\cdot 10^{15}$ eV$^2$.

The equations of motion derived from (\ref{lagrangian})  combined with the Jacobi identity $\pa_{\left[\alpha\right.}F_{\left.\beta\gamma\right]}=0$ are
\beqar
&\displaystyle\nonumber
\nabla \bm{H} = 0,\quad
\nabla\times \bm{E} = -\dot{ \bm{H}},\quad
\nabla \bm{E} = -g(\bm{H}\nabla a),
&\\&\displaystyle\label{AxMaxwell}
\nabla\times \bm{H} = \dot{ \bm{E}}+g\Bigl(\bm{H} \dot a -\bm{E}\times\nabla a\Bigr),\qquad
&\\&\displaystyle\nonumber
\ddot a - \nabla^2 a + m_a^2 a =g(\bm{E}\bm{H}).
&
\eeqar
Here the dot means the differentiation with respect to time.
For the real physical fields we can take $g$ as a small parameter and expand Eqs.~(\ref{AxMaxwell}) in powers of $g$. In zeroth order the equations for the axions and the electromagnetic field are separated. The galactic halo axions are nonrelativistic and  their energy on the Earth is close their rest energy. The relative velocity between the Earth and the galactic center $\beta\sim 10^{-3}$ and an axion velocity spread has a conservative estimation of the same order so that the axion energy is $E_a\approx m_a(1+\frac{1}{2}\beta^2)$. The corresponding de Broglie wavelength $\lambda=2\pi/\beta m_a$  is much greater then a detector size, and hence a spatial dependence in axion dynamics is negligible. Then the first-order equations are
\begin{subequations}
\label{Ax 1order}
\beqar
&\displaystyle\label{Ax 1order div}
\nabla \bm{\epsilon} = 0,\qquad
\nabla \bm{h} = 0,\qquad
&\\&\displaystyle\label{Ax 1order rot}
\nabla\times \bm{\epsilon} = -\dot{ \bm{h}},\qquad
\nabla\times \bm{h} = \dot{ \bm{\epsilon}}+g\bm{H}_0 \dot a,\qquad
&
\eeqar
\end{subequations}
where $a=A\cos m_at$ and $\bm{H}_0$ is a large static magnetic field.

It is clear from Eqs.~(\ref{Ax 1order}) that just an \emph{inhomogeneous} magnetic field is perturbed  by the axions. For the point-like JJs the magnetic field effect is negligible \cite{Likharev} and the axions effectively drop out of the consideration. In the context of the conventional axion-photon interaction the axions is able to reveal themselves in the extended JJs and SQUIDs whose current-voltage (\IV) characteristics depend on a magnetic flux containing in the device. According to Eqs.~(\ref{Ax 1order}) the axions interacting with the \emph{transverse} magnetic field $\bm{H}_0=H_0(x)\bm{e}_y$ produce the periodic \emph{longitudinal} field $\bm{h}=h(x,t)\bm{e}_z$. A SQUID ring in the $xy$-plane ensures that the magnetic flux threading the SQUID pickup loop is caused by the only contribution $\bm{h}$ from the axion-phonon interaction.

\begin{figure}[t]
\begin{center}\includegraphics[width=0.5\textwidth]{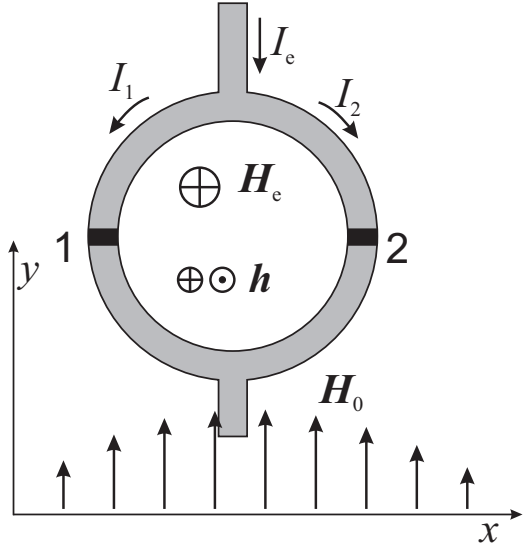}\end{center}%
\caption{A conceptual sketch of the proposed experiment. The static magnetic field $\bm{H}_0$  is inhomogeneous along the $x$-axis. The axion-induced perturbations oscillate in the direction perpendicular to the figure plane.}\label{fig-SQUID}
\end{figure}

A suitable description of a dc SQUID response is given in the framework of the resistively and capacitively shunted junction (RCSJ) model \cite{Likharev,Barone}. For simplicity we consider two identical JJs incorporated into the SQUID ring (see Fig.~\ref{fig-SQUID})

According the RCSJ model a bias current $I_\text{e}$ entering the SQUID loop splits into two parts
\beq
I_\text{e}=I_1+I_2,
\eeq
where the currents through the junctions are
\beq\label{dcSQUIDeqs gen}
I_k=C \dot V_k +\frac{V_k}{R}+I_\text{c}\sin\varphi_k,\qquad
k=1,2.
\eeq
Here $\varphi_k$ are the phase differences of the junctions. The voltages $V_k$ across the junctions evolve according to the Josephson relation
\beq\label{Josephson2}
V_k=\frac{1}{2e} \dot\varphi_k.
\eeq
The capacity $C$, the resistance $R$, and the critical current $I_\text{c}$ are the same for both junctions.

The phase differences $\varphi_1$ and $\varphi_2$ are related to a total magnetic flux $\Phi$ through the pickup loop by
\beq
\varphi_1-\varphi_2=2\pi\frac{\Phi}{\Phi_0},\qquad
\eeq
where $\Phi_0=\pi/e\approx 2.07\times 10^{-15}$ Wb is the magnetic flux quantum. The total flux includes contributions from an applied external magnetic field and from the currents $I_1$ and $I_2$,
\beq
\Phi=\Phi_\text{e}-L(I_1-I_2).
\eeq

Using notations
\beq\label{phichi}
\frac{\varphi_1-\varphi_2}{2}=\varphi, \qquad \frac{\varphi_1+\varphi_2}{2}=\chi,
\eeq
one easily obtains a pair of dimensionless equations
\beqar
&&\label{dlEq1}
\beta_\text{L}^{-1}(\varphi-\varphi_\text{e}) + \beta_\text{c}\ddot\varphi + \dot\varphi +\sin\varphi\cos\chi=0,
\\&&\label{dlEq2}
\beta_\text{c}\ddot\chi + \dot\chi +\sin\chi\cos\varphi=i,
\eeqar
where $\beta_\text{c}$ is the Stewart-McCumber parameter, which is equal to the ratio of the squares of the characteristic frequency $\omega_\text{c}=2eI_\text{c}R$ to the plasma frequency $\omega_\text{p}=(2eI_\text{c}/C)^{1/2}$ of the junction,  $\beta_\text{L}=2\pi LI_\text{c}/\Phi_0$ is the screening parameter, $\varphi_\text{e}=\pi\Phi/\Phi_0$ is the normalized applied flux, and $i=I_\text{e}/2I_\text{c}$ is the normalized bias current. The dot now means the differentiation with respect to dimensionless time $\tau=\omega_\text{c}t$.

For the sake of simplicity we restrict our consideration to the negligible junction capacitance ($\beta_\text{c}\ll 1$) and SQUID inductance ($\beta_\text{L}\ll 1$). These constraints put the SQUID in non-hysteretic regimes. The former corresponds to the strongly overdumped limit and helps to avoid hysteresis in the \IV curve while the latter helps to avoid magnetic hysteresis \cite{Tinkham}. Besides, the latter constraint strictly relates the difference between the JJs phases to the applied magnetic flux. On the one hand these conditions considerably simplify our consideration, and on the other hand they are realized in many practical applications. Under these approximations the Eq.~(\ref{dlEq1}) reduces to $\varphi=\varphi_\text{e}$ and the Eq.~(\ref{dlEq2}) becomes
\beq\label{dcSQUIDeq}
\dot\chi +\sin\chi\cos\varphi_\text{e}=i.
\eeq
The Eq.~(\ref{dcSQUIDeq}) implies that the ds SQUID behaves as a single Josephson junction with the critical current depending on the applied magnetic flux. For the constant flux the exact solution of Eq.~(\ref{dcSQUIDeq}) is $\chi=\text{const}$ when $i_0<|\cos\varphi_\text{e}|$, and
\beq\label{chi0}
\chi = 2\arctan \left[\sqrt{\frac{i_0+\cos\varphi_\text{e}}{i_0-\cos\varphi_\text{e}}} \tan\frac{v \tau}{2}\right]-\frac{\pi}{2},
\eeq
where
\beq\label{SQUIDvoltage}
v=\sqrt{i_0^2-\cos^2\varphi_\text{e}},
\eeq
otherwise \cite{Likharev}. The subscript 0 is used here to fix the zero-order solution that forms a smooth curve in the \IV characteristic of the SQUID. According to Eqs.~(\ref{Josephson2}) and (\ref{phichi}) the time average of $\dot\chi$  corresponds to a normalized voltage measured experimentally. For  solution~(\ref{chi0}) the average yields $\langle\dot\chi\rangle=v$ \cite{Likharev,Barone}.

A noticeable feature of JJs is the occurrence of current steps (so-called Shapiro steps) in the \IV curves when one applies an additional ac current \cite{Likharev}. The similar steps arise in the \IV characteristic of the SQUID if the applied magnetic flux has a sinusoidal contribution. To demonstrate this effect we split the normalized magnetic flux $\varphi_\text{e}$ into constant and small periodic contributions
\beq
\varphi_\text{e} = \phi_0 + \phi_1\sin v \tau,\qquad \phi_1\ll\phi_0.
\eeq
In this case a first-order term is added on the right hand side of Eq.~(\ref{dcSQUIDeq})
\beq\label{dcSQUIDaxEq}
\dot\chi +\sin\chi\cos\phi_0=i +\phi_1\sin\phi_0\sin\chi\sin v \tau.
\eeq
In our consideration $\phi_0$ is set up by an external constant field $\bm{H}_\text{e}=H_\text{e}\bm{e}_z$ whereas the periodic perturbation is meant to be generated by the DM axions.

There is no need to directly solve Eq.~(\ref{dcSQUIDaxEq}) to find out the modifications produced by the periodic flux in the \IV characteristic. For our purpose solution (\ref{chi0}) with an arbitrary function $\theta(\tau)$ instead the phase $v \tau$ is to be substituted into Eq.~(\ref{dcSQUIDaxEq}). The next step is the time average to made rapidly oscillating terms to vanish:
\beq
v\dot\theta-v^2=(i-i_0)i_0+\phi_1\sin\phi_0\langle\cos(\theta-v \tau)\rangle.
\eeq
Here the angle brackets denote the time average. A natural substitution $\Theta=\theta-v \tau$ leads to the equation
\beq\label{ShStepEq}
\dot\Theta + \frac{\phi_1\sin\phi_0}{2v}\sin\Theta = \frac{i-i_0}{v},
\eeq
which exactly coincides with Eq.~(\ref{dcSQUIDeq}). As discussed above the solution of Eq.~(\ref{ShStepEq}) is  $\Theta=\text{const}$ for  $|i-i_0|<\frac{1}{2}\phi_1|\sin\phi_0|$ implying a constant voltage for some current range. It describes a step in the \IV curve at the voltage $V$ corresponding to the axion energy $m$ according to $m_a=eV=evRI_\text{c}$. A normalized height of this step
\beq\label{stepH1}
\Delta i= \phi_1|\sin\phi_0|
\eeq
is proportional to the longitudinal distortion of the magnetic field induced by the axions. The distortion arises from the transverse inhomogeneous field $\bm{H}_0$ according to the equation
\beq\label{long h}
\ddot h - \nabla^2 h = g |\nabla\times\bm{H}_0|\dot a,
\eeq
following immediately from Eqs.~(\ref{Ax 1order rot}).
The constant field $\bm{H}_e$  is naturally excluded from Eq.~(\ref{long h}) and therefore it only modulates signal (\ref{stepH1}). The maximum value of the current step corresponds to $\phi_0=\pi(n+\frac{1}{2})$ with an integer $n$. In this case the phases of the JJs are opposite, i.e. $\varphi_1-\varphi_2=\pm\pi$, and the supercurrents flow in opposite directions.

\begin{figure}[t]
\begin{center}\includegraphics[width=0.5\textwidth]{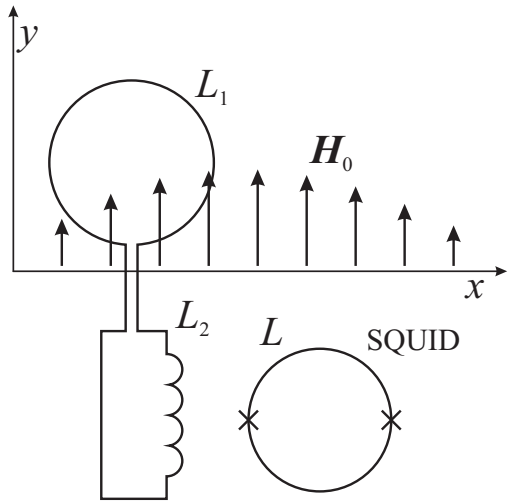}\end{center}%
\caption{A schematic sketch of the detector. It uses the flux transformer with a pickup loop $L_1$ and an input coil $L_2$ to isolate the SQUID from the static magnetic field $\bm{H}_0$.}\label{fig-Detector}
\end{figure}

A schematic diagram of the proposed axion detector is shown in Fig.~\ref{fig-Detector}.
The required field $\bm{H}_0$ can be produced, for example, in the TM$_{mn0}$ mode of a rectangular microwave cavity. In this case the solution of Eq.~(\ref{long h}) with proper boundary conditions is a sum of natural modes with frequencies depending on a cavity cross-section. By varying its sizes the cavity can be tuned to the axion field frequency $m$. As it is described by Eq.~(\ref{long h}), the corresponding mode becomes dominant and its amplitude linear increases with time while damping is not taken into account.  For a real cavity the increasing factor $mt$ is substituted by an appropriate quality factor $Q$ and the resonant mode amplitude becomes $h_\text{res}=\frac{1}{2}gAQH_0$.

This microwave mode can be detected by the dc SQUID if the half-wavelength of this mode is greater than the pickup loop. Because of the need to isolate the SQUID from the strong magnetic field $\bm{H}_0$, the proposed scheme involves a flux transformer. It includes a pickup loop of inductance $L_1$ and an input coil of inductance $L_2$ coupled to the SQUID via a mutual inductance $M=k\sqrt{2LL_2}$, and the SQUID is enclosed in a superconducting shield. The magnetic flux $\Phi_\text{a}=hS$ caused by the axions induces a current $I_\text{t}$ in the transformer circuit according to $\Phi_\text{a}=I_\text{t}(L_1+L_2)$. Then a flux in the SQUID is
\beq
\Phi_1=MI_\text{t}=\frac{k\sqrt{2LL_2}}{L_1+L_2}=\mu \Phi_\text{a}.
\eeq

A response is a steplike signal in the \IV curve at the voltage corresponding to the axion mass $V=m_a/e$ with the normalized current amplitude
\beq
\Delta i\approx \frac{\mu\alpha g_\gamma}{\pi}\cdot\frac{\sqrt{2\rho_a}}{mf_a}\cdot\frac{H_0 SQ}{\Phi_0},
\eeq
where $\rho_a=\frac{1}{2}m_a^2A^2$ is the axionic DM energy density and $S$ is the pickup loop area.

The total quality factor includes two contributions due to absorption into the cavity walls $Q_\text{c}$, and due to the axion energy spread $Q_a$
\beq
\frac{1}{Q} = \frac{1}{Q_\text{c}} + \frac{1}{Q_a}.
\eeq
Using the superconducting cavity the $Q_\text{c}$ factor can be brought to $10^{10}$ \cite{Padamsee} whereas the $Q_a$ factor is in inverse proportion to the axion energy dispersion and is model-dependent. Within the isothermal sphere model  of DM halos the velocity spread is of order of the circular velocity $\beta\sim 10^{-3}$ and so $Q_a\approx \beta^{-2}\sim 10^{6}$. In this case the total quality factor $Q\approx Q_a$. A different estimation is supported by a model developing an idea that the axions form a Bose-Einstein condensate \cite{Sikivie2}. In this state the velocity dispersion $\delta\beta\sim 10^{-7}$ (some authors \cite{Mielczarek} suppose the considerably lower estimation $\delta\beta\sim 10^{-12}$) and the corresponding quality factor $Q_a\approx(\beta\delta\beta)^{-1}\sim 10^{10}$ has the same order as $Q_\text{c}$.

Taking into account that the local galactic halo DM energy density near the Earth is estimated as $\rho_\text{DM}\approx 0.3$ GeV/cm$^3$ \cite{comment2} we have
\beqar
\Delta i &\sim & 10^{-4}\left(\frac{\rho_a}{\rho_\text{DM}}\right)^{1/2} \left(\frac{\rho_\text{DM}}{\text{GeV}/\text{cm}^3}\right)^{1/2}
\!\!\times\\ \nonumber && \times%
\left(\frac{H_0}{\text{T}}\right) \left(\frac{S}{\text{cm}^2}\right) \left(\frac{Q}{10^6}\right).
\eeqar

The signal amplitude  is directly independent of the axion mass. It is conceivable that this makes possible to advance the proposed method for a wide mass range.
However, constraints on the axion mass arise from the size of the real device. To detect the signal the pickup loop ought to fit in the half-wavelength corresponding to the axion mass so the proper masses is restricted by $m_a\lesssim 10^{-4}$~eV.

A sensitivity of up-to-date magnetometers with a flux transformer is about $10^{-13}$ T and their noise level is of order $10^{-16}$ T$\cdot$Hz$^{-1/2}$. This level is quite acceptable for the proposed experimental strategy. However, a clearness of the scheme is primarily conditioned by thermal fluctuations in the SQUID. The fluctuations smoothes the current step as well as the \IV curve as a whole. A description of the fluctuations involves additional stochastic currents $I_{\text{F}1,2}$ in the r.h.s. of Eqs.~(\ref{dcSQUIDeqs gen}), which became Langevin equations. The currents $I_{\text{F}k}$ are considered as a white noise and stochastic methods, such as the Fokker-Plank equations one, are employed. This subject is beyond the scope of the present article and will be investigated further. The qualitative estimation of the thermal fluctuations effect is based on the parameter $\gamma=2eT/I_\text{c}$, the ratio of the thermal energy to the supercurrent energy. It describes the intensity of the fluctuations in the JJs and smoothing of the \IV curve. Smoothing of the current step is described by the effective noise parameter
\beq
\gamma_\text{eff}= \gamma\,\frac{2i_0^2+\cos^2\phi_0}{2i_0v\phi_1\sin\phi_0}
\eeq
arising from the transformations from Eq.~(\ref{dcSQUIDaxEq}) to Eq.~(\ref{ShStepEq}). The small normalized magnetic flux is in the denominator, so that $\gamma_\text{eff}>\gamma$ and smoothing of the current step takes place at smaller fluctuations than that of the \IV characteristic as a whole. The similar behavior is also inherent for the classical Shapiro steps in the ordinary JJ \cite{Likharev}.

In conclusion it makes sense to summarize features of the proposed experimental technique. It provides for the microwave cavity where the DM axions bring about the transformations from the transverse magnetic field energy to the longitudinal perturbation energy. It is evident that to use cavity is not only way to obtain this sort of perturbations although an undoubted advantage of the cavity is its high quality factor. In such an approach the cavity serves as a transformer and as an amplifier at the same time. The amplified perturbations oscillate with the frequency corresponding  to the axion mass. To detect this mode the frequency ought to be synchronized with the voltage across the dc SQUID. By this means the SQUID acts not like a magnetometer but like a frequency-to-voltage converter. If the DM is entirely axionic or, at least the axions constitute a considerable fraction of the DM, the signal is found to be small but detectable.

This work was supported by the Russian Foundation for Basic Research through Grant No.~14-02-00598.

\end{document}